\documentclass{aa}
\usepackage{graphicx}

\usepackage{psfig}
\usepackage{epsfig}
\usepackage{txfonts}
\begin{document}

\title{Do weak magnetic fields prevent hydrogen from accreting onto  
metal-line white dwarf stars? \thanks{Based on observations 
collected at the European Southern Observatory, Paranal, Chile (ESO Programme 
66.D-0541)}  }

\author{Susanne Friedrich\inst1
   \and
   Stefan Jordan\inst2\inst3
   \and 
   Detlev Koester\inst4
}

\offprints{S. Friedrich}

\institute{Max-Planck-Institut f\"ur extraterrestrische Physik,
  Giessenbachstr., 85748 Garching, Germany, \email{sfriedrich@mpe.mpg.de}
\and
   Institut f\"ur Astronomie und Astrophysik, Sand 1, 72076 T\"ubingen,
   Germany, 
\and
   Astronomisches Rechen-Institut, M\"onchhofstr. 12-14, 
   69221 Heidelberg, Germany, \email{jordan@ari.uni-heidelberg.de}
  \and
   Institut f\"ur Theoretische Physik und Astrophysik, 
             Universit\"at Kiel, 24098 Kiel, Germany,
             \email{koester@astrophysik.uni-kiel.de}
}


\abstract{
The widely accepted assumption is that metals detected in the spectra of 
a few cool helium-rich white dwarfs cannot be of primordial origin and 
therefore must be accreted from the interstellar medium. However, 
the observed abundances of hydrogen are much too low to be compatible 
with the high accretion rates inferred from metal accretion. 
Hydrogen accretion is therefore suppressed compared to metal 
accretion. The hypothesis most widely discussed as cause for this 
``hydrogen screening'' is the propeller mechanism: Metals are accreted 
in the form of grains onto a slowly rotating, weakly magnetized white 
dwarf, whereas ionized hydrogen is repelled at the Alfv\'en radius. 
We have obtained circular polarization spectra of the helium-rich white dwarfs 
\object{GD\,40} (\object{WD0300-013}) and 
\object{L745-46A} (\object{WD0738-172)} --- 
which both show strong metal lines as well as
hydrogen --- in order to search for signatures of a weak magnetic field.
The magnetic field strengths necessary for the propeller mechanism to work in
these stars are at least 144000~G and 3000~G, respectively. Whereas L745-46A
might have a magnetic field of about $-$6900~G no magnetic field could be 
found with an upper limit for the field strength of 4000~G (with 99\%
confidence) for GD\,40.

}

\authorrunning{S. Friedrich et al.}

\titlerunning{Do magnetic fields prevent hydrogen from accreting onto  
 metal-line white dwarfs?}

\maketitle

\section{Introduction}
Some of the helium-rich white dwarfs at the cool end of the white dwarf
sequence show evidence for metal lines in their spectra. Since radiation 
forces are not strong enough to compete with gravity for temperatures below
40000~K it is expected that metals sink down on time scales which are
short compared to the cooling age. If nevertheless metals are 
observed in the atmospheres, as in the DZ and related spectral types, 
they must have come from outside the star. The most popular mechanism 
for this is accretion from interstellar matter.

Based on earlier studies (Koester \cite{koester76}; Wesemael 
\cite{wesemael79}; Vauclair et al. \cite{vauclair79}) 
concerning accretion processes onto cool white dwarfs, Dupuis et 
al. (\cite{dupuis92}, \cite{dupuis93a}, \cite{dupuis93b}) 
developed a theoretical model which 
predicts element abundances in cool white dwarf atmospheres 
under the assumption that elements are accreted in solar element 
proportions from the interstellar medium. They followed the 
suggestion of Wesemael (\cite{wesemael79}) that due to the patchy structure 
of the interstellar medium a white dwarf may experience highly variable 
accretion rates during its life. Most of the time (about $5\cdot 10^7$ 
years assumed in the strongly idealized scenario) the white dwarf will 
accrete at very low rates ($5\cdot 10^{-20}$ M$_{\sun}$/yr) which lead 
to metal abundances in the photosphere well below the detection limit. 
Only during (or shortly after) encounters with denser patches in the 
interstellar medium (typical density 10 particles cm$^{-3}$ and 
temperature 100~K), metals are accreted with rates high enough to 
lead to detectable features in the spectra.  

Whereas this scenario can account for the observed metal abundances 
in helium-rich white dwarfs, the observed abundances of hydrogen 
are much too low to be compatible with the accretion rates inferred 
from metal accretion (Dupuis et al. \cite{dupuis93b}). 
In a recent study Wolff et al. (\cite{wolff02}) analyzed the UV 
spectra of ten DZ and DBZ stars which form a homogenous data set and 
concluded that the observed hydrogen abundance is more than two orders
of magnitude to low. Furthermore DZ white dwarfs are not found in or 
close to known dense clouds (Aannestad et
al. \cite{aannestad93}). One possible explanation is that hydrogen 
accretion is reduced 
compared to metal accretion. The mechanism most widely discussed as 
reason for this ``hydrogen screening'' is the propeller mechanism 
of Illarionov \& Sunyaev (\cite{illarionov75}). 

\section{The propeller mechanism}
The propeller mechanism was originally developed for accretion onto 
neutron stars (Illarionov \& Sunyaev 1975) and adopted to white dwarfs by
Wesemael \& Truran (\cite{wesemael82}). Wesemael \& Truran proposed
that metals are accreted in the form of grains onto a slowly rotating,
weakly magnetized white dwarf, whereas ionized hydrogen is repelled at
the Alfv\'en radius. 
Fig.~1 in Wesemael \& Truran shows the domain of allowed
Alfv\'en radii as a function of luminosity (or temperature) of the central
star. The domain is restricted to higher temperatures by setting the dust 
evaporation time scale at the Alfv\'en radius equal to the free-fall time from
the accretion radius, to lower temperatures by equating the ionization 
radius to the Alfv\'en radius. For typical white dwarfs ($M=0.6M_\odot$) 
and accretion parameters ($\dot{M}=10^{-15}M_\odot$/yr, $v=50$km/s) this 
region is roughly between $T_{\hbox{eff}}=7500$~K and 
$T_{\hbox{eff}}=15000$~K. The lower boundary can be bypassed if accretion 
from a warm ISM phase with temperature $T\approx 10000$~K and density 
$n\approx 1$~cm$^{-3}$ is assumed. In
such a phase, the hydrogen would be ionized, but the metals may still be
partially locked up in grains (Alcock \& Illarionov \cite{alcock80}).
Given the effective temperature of a star
the minimum Alfv\'en radius, which in turn corresponds to a minimum
magnetic field strength necessary to let the propeller mechanism work, 
can be calculated. 

To our knowledge it has never been tested whether in helium-rich 
metal-line white dwarfs the necessary conditions for 
rotation, magnetic field, and strength of hydrogen-ionizing radiation
are indeed fulfilled to make this mechanism operate over the entire range 
of $T_{\hbox{eff}}$ from 7500~K to 15000~K. 

\section{Observation and data reduction}
With effective temperatures of 15150~K and 7500~K, respectively, 
GD\,40 (WD0300-013, V=15.56) and L745-46A (WD0738-172, V=13.04) 
mark the upper and lower boundary of the temperature regime of 
the propeller mechanism. Both stars are well studied and atmospheric
parameters as well as element abundances are known with sufficient accuracy. 
Based on the accretion-diffusion model of Dupuis et al. (\cite{dupuis93b}) 
the hydrogen accretion rates necessary to provide the observed hydrogen
abundances are about three or four orders of magnitude lower than the average
accretion rate which is assumed to account for the observed metal abundances 
(Friedrich et al. \cite{friedrich99}; Koester \& Wolff \cite{koester00}).
Therefore, hydrogen accretion must be efficiently reduced even onto L745-46A
which shows one of the highest hydrogen abundances observed in cool
helium-rich white dwarfs.  

For the above reasons both stars are ideal candidates to search for magnetic 
fields driving the propeller mechanism. The VLT with FORS1 and 
PMOS offers for the first time the possibility to obtain
circular polarization spectra with high enough signal-to-noise ratios at 
acceptable exposure times to search for these magnetic fields as it has been 
demonstrated by Aznar Cuadrado et al. (\cite{aznar04}) for white dwarfs.

Flux and circular polarization spectra were obtained with the
VLT UT1 and PMOS in December 2000 in service mode. Grism 600R (dispersion 
1\AA /pix) was used with order sorting filter GG435 resulting 
in a spectral range of about 5300-7350\AA . Several individual 
exposures were taken for each star.  
After each exposure the retarder plate was rotated by 90 degrees in order to
account for improper flat field corrections, instrumental polarizations and
depolarization. The resulting total exposure times for GD\,40 and L745-46A 
were 3.3 hours and 2.7 hours. The signal-to-noise ratio in the 
vicinity of H$\alpha$ and the helium lines of GD\,40 amounts to 
190 and 260, respectively, and to 280 for H$\alpha$ of L745-46A. 
Data reduction was done with 
standard IRAF routines. Circular polarization was first calculated
for each individual retarder position from the difference of the double 
spectrum, produced by the Wollaston prism, divided by the sum of the 
double spectrum. Polarization spectra, corrected for instrumental 
influences, were then derived from the difference of two consecutive 
circular polarization spectra with different quarter wave plate positions.
Subsequently these polarization spectra were averaged to
get a mean circular polarization spectrum. The flux spectrum is
given by the sum of both spectra. All individual flux spectra were
averaged to get a mean flux spectrum.

In addition a star in the field of view of L745-46A, which we assume to be 
unpolarized, was observed. It served as a supplementary check for 
instrumental introduced polarization. The mean circular polarization in a
spectral range between 5500\AA\ and 6800\AA\ amounts
to 0.01\%, 0.02\%, and 0.03\% for the field star, GD\,40, and 
L745-46A, respectively, one order of magnitude lower than the expected 
values for kilogauss magnetic fields.  

\section{Determination of magnetic fields}
\subsection{Weak-field approximation}
According to the theory of line formation in a weak magnetic field the splitting 
of a spectral line is proportional to the mean longitudinal
field $B_{\rm l}$, i.e. the component of the magnetic field 
along the line of site averaged over the visible stellar surface.
Provided that the Zeeman splitting of a spectral line is small
compared to intrinsic (thermal and pressure) broadening (e.g. Angel \&
Landstreet \cite{angel70}) the amount of polarization can be determined 
by the weak-field approximation, which is valid even in the presence of 
instrumental broadening, but is not generally correct in
the case of rotationally broadening (Landstreet \cite{landstreet82}):

$$\frac{V}{I}=- g_{\rm eff} \frac{e}{4 \pi m_e c^2} \lambda^2
\frac{dI}{I_\lambda d\lambda}B_{\rm l}$$
with $e/4 \pi m_e c^2 \approx 4.67 10^{-13}$ \AA$^{-1}$ G$^{-1}$, 
and $g_{\rm eff}$ the effective Land\'e factor which equals 1 for hydrogen
Balmer lines (e.g. Bagnulo et al. \cite{bagnulo02} and citations therein).
For the \ion{He}{i} line at 5875\AA\ $g_{\rm eff} =1.16$ (Leone et al. 
\cite{leone00}).

In our case the weak-field approximation is justified because intrinsic
broadening is much larger than the separation of line components due to 
Zeeman splitting which is of the order of 0.1\AA\ for field strengths of 
10 kG. For higher field strengths which are called for by the 
propeller mechanism for GD\,40 the intrinsic broadening is smaller than the
magnetic broadening and the weak-field approximation reduces to a
order-of-magnitude estimation (Landstreet \cite{landstreet82}). 

Due to the lower signal-to-noise ratio at the H$\alpha$ line of GD\,40 
the synthetic flux spectrum $I(\lambda)$ from Section 5.1.1. was used 
to calculate ${V}/{I}$. For L745-46A the S/N-ratio was sufficiently high
to use the observed flux spectrum.
${dI}/{d\lambda}$ is approximated by the average of 
$(I_{i+1} - I_{i})/(\lambda_{i+1}-\lambda_{i})$ and 
$(I_{i} - I_{i-1})/(\lambda_{i}-\lambda_{i-1})$ for a given wavelength
$\lambda_{i}$. 

\subsection{Fitting of the magnetic field}

In order to determine the mean longitudinal component of the magnetic field
the observed circular polarization was compared to the predictions of the
weak-field approximation in an interval of $\pm$20\AA\ around H$\alpha$ and
the \ion{He}{i} lines at 5875\AA\ and 6678\AA . The best-fit for $B_{\rm l}$, the only
free parameter, was found by a $\chi^2$ minimization procedure. If 
no magnetic field were present, all deviations from zero would be due to
noise. This can be expressed by the standard deviation over the respective
intervals around the respective lines. If the best-fit magnetic field is close
to zero the reduced $\chi^2$ should be close to unity. However, if a 
magnetic field
is present some deviations from zero are not due to noise and the reduced
$\chi^2$ is smaller than 1. Following Press et al. (\cite{press86}) the
statistical error is determined from the rms deviation of the observed
circular polarization from the best-fit model. The 1$\sigma$ (68.3\%)
confidence range for a degree of freedom of 1 is the interval of $B_{\rm l}$
where the deviation from the minimum is $\Delta\chi^2=1$ (6.63 for the 99\%
confidence range). This is a purely
statistical error and does not account for systematic errors. For details see
Aznar Cuadrado et al. (\cite{aznar04}).

\subsection{Magnetic field strengths for the propeller mechanism}
Minimum magnetic field strengths necessary for the propeller mechanism to work
were calculated according to Wesemael \& Truran (\cite{wesemael82}). The
logarithm of the Alv\'en radius was estimated from their Fig.~1.  
Assuming a white dwarf mass of $0.6M_{\odot}$, a radius of
$0.013 R_{\odot}$, an accretion rate of $10^{-15} M_{\odot}$/yr the 
corresponding minimum magnetic field strengths from their Eq.~4a are 
144000$^{+25000}_{-17000}$~G and 3000$\pm 500$~G for GD\,40 and L745-46A, 
respectively. The error accounts for a reading error of 1mm in their Fig.~1. 
Further errors are introduced by the stellar radius which enters the formula
with third power and the accretion rate. Lower radii and higher accretion 
rates would increase the minimum magnetic field strength.

\section{Results}
\subsection{GD\,40}
\subsubsection{Hydrogen abundance}

\begin{figure}[htbp]
\includegraphics[width=0.45\textwidth]{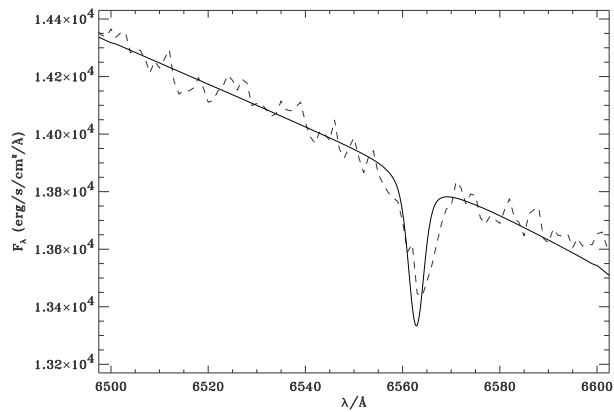}
\caption[]{Observed flux spectrum (dashed) of GD\,40 at the position of 
H$\alpha$ with a non-magnetic model 
spectrum (solid, T$_{\mbox{eff}}=15150$, $\log g=8$) over-plotted. 
The hydrogen abundance needed to reproduce the line profile of 
H$\alpha$ amounts to $5\cdot10^{-7}$ relative to helium} 
\label{gdflux}    
\end{figure}

The flux spectrum of GD\,40 clearly shows an H$\alpha$ absorption line which 
confirms the detection of this line by Greenstein \& Liebert 
(\cite{greenstein90}, Fig.\,\ref{gdflux}). In disagreement to these 
authors, who determined an equivalent width of 1\AA , we measured 0.2\AA\ 
only. We attribute the difference in the equivalent widths to the much 
better quality of our spectra compared to those of Greenstein \& Liebert.
As a consequence the hydrogen abundance turned out to be a factor of 10 
below the value previously determined (Friedrich et al. \cite{friedrich99}). 
In order to 
estimate the hydrogen abundance, we took the best fit, non-magnetic 
model atmosphere with $T_{\mbox{eff}}=15150$, $\log g=8$ from Friedrich et 
al. (\cite{friedrich99}) and varied the hydrogen abundance while keeping 
all other 
element abundances and the atmospheric parameters fixed. This resulted in a 
new hydrogen abundance of $5\cdot10^{-7}$ relative to helium for GD\,40.

\subsubsection{Circular polarization}

\begin{figure}[htbp]
\includegraphics[width=0.45\textwidth]{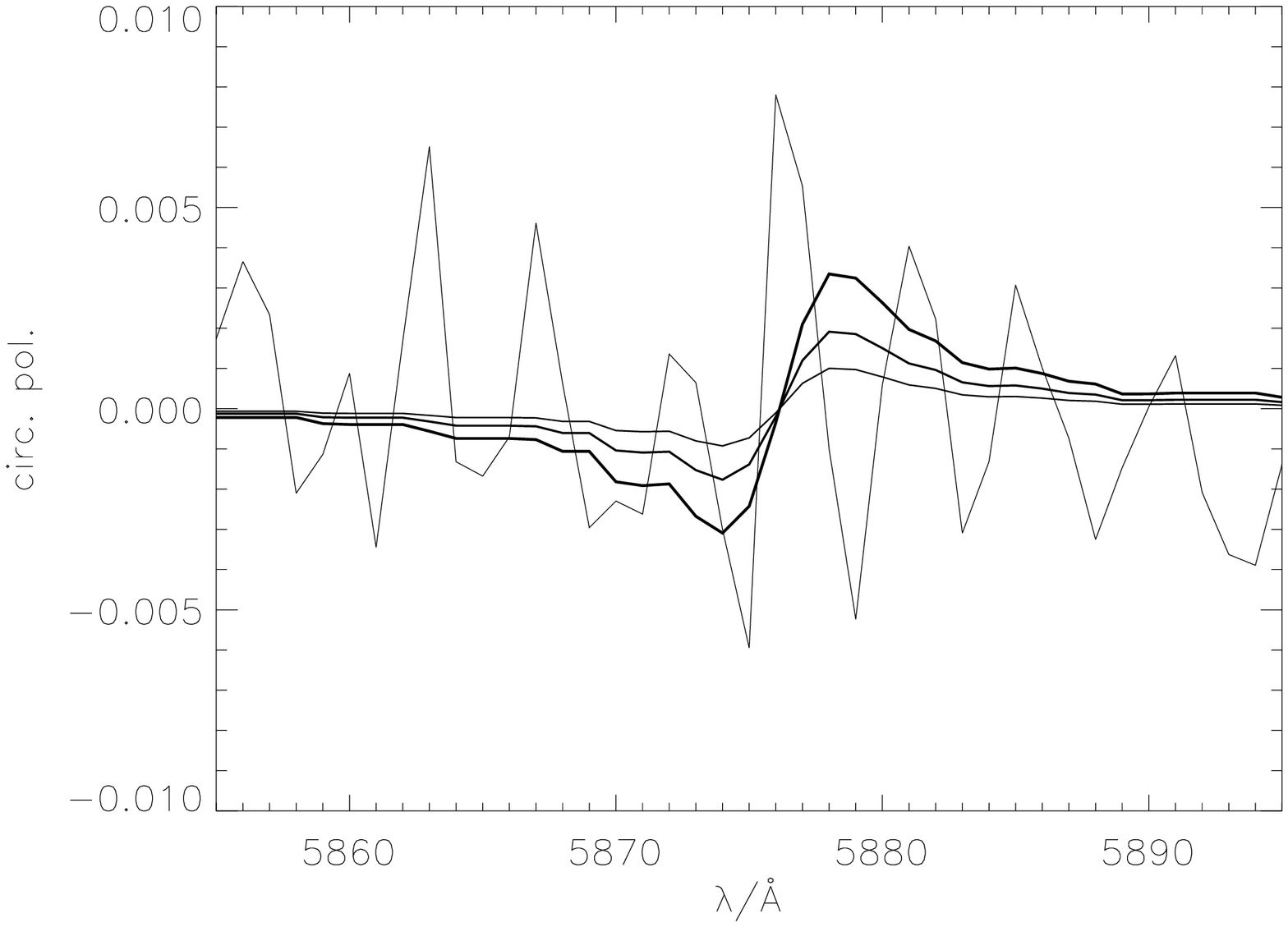}
\includegraphics[width=0.45\textwidth]{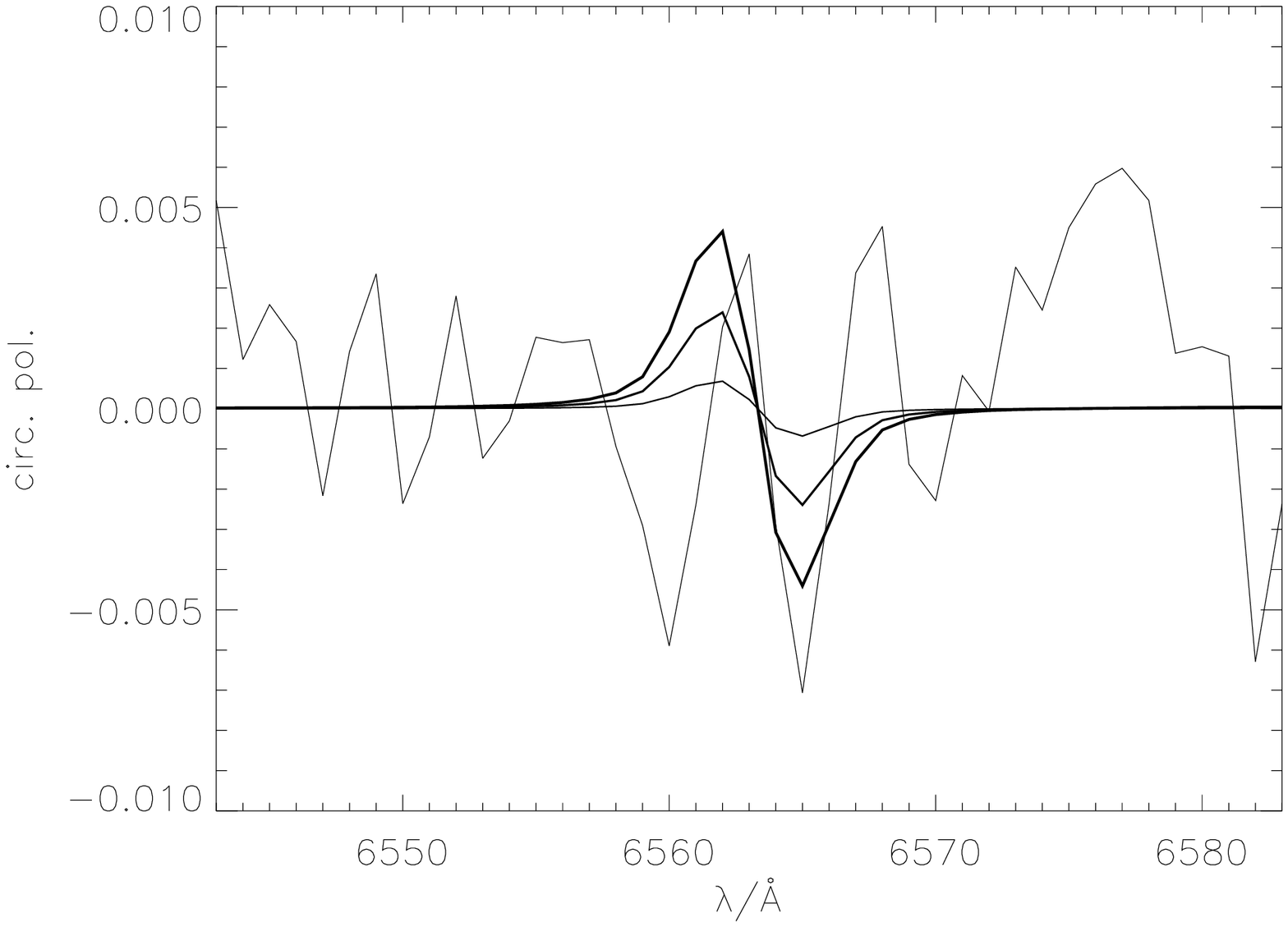}
\includegraphics[width=0.45\textwidth]{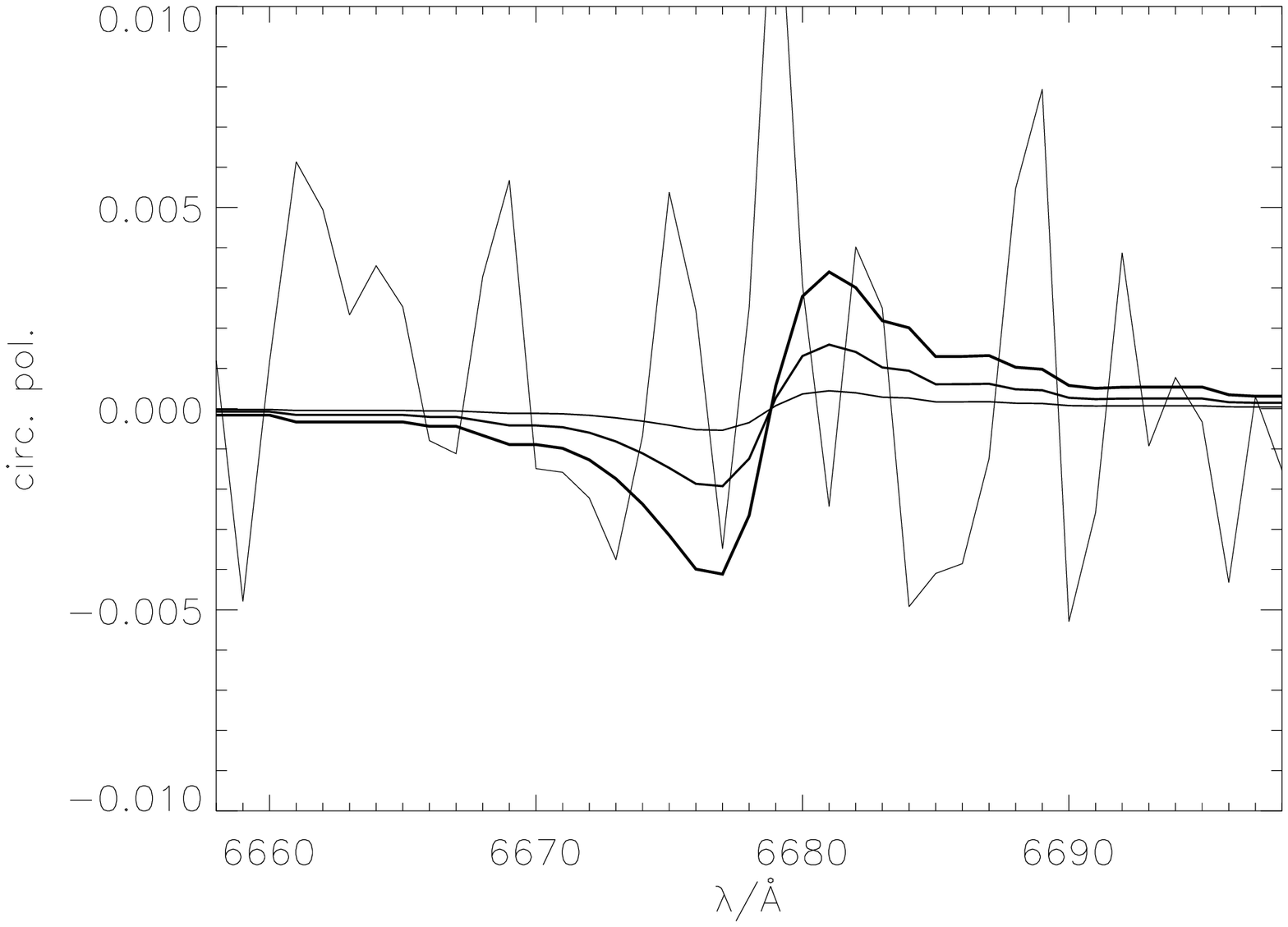}
\caption[]{Observed (gray) circular polarization spectrum of GD\,40 
  together with the
  respective polarization spectra (bold) predicted by the weak-field
  approximation for field strengths of 2747~G, 9639~G, and 17747~G for
  H$\alpha$ (top), $-1683$~G, $-3211$~G, and $-5619$~G for the \ion{He}{i} 
  line at 5875\AA\ (middle), and $-716$~G, $-2564$~G, and $-5474$~G for 
  the \ion{He}{i} line at 6678\AA\ (bottom). For each spectral line 
  the first field
  strength denotes the best fit value, the second and third value the 
  best fit value plus the 68.3\%  and 99\% confidence range, respectively} 
\label{gd_pol}    
\end{figure}

The inspection of the individual polarization spectra of GD\,40 did not show
any obvious features or variations in the polarization which cannot be 
attributed to noise. The magnetic field strengths calculated with the 
theoretical weak-field approximation are below the 1$\sigma$ error. In 
addition different magnetic field orientations are found (indicated by 
positive and negative signs of $B_{\rm l}$) for H$\alpha$ and the 
two \ion{He}{i} lines. For the H$\alpha$ line the field
strength and the respective 68.3\%/99\% confidence ranges amount to 
($2747\pm 6892/15000$)~G, and for the two \ion{He}{i} lines at 
5875\AA\ and 6678\AA\ to ($-1683\pm 1528/3936$)~G and ($-716\pm 1848/4758$)~G, 
respectively. A fit to all three lines
results in a magnetic field strength of 1131~G and a 99\% confidence level of
$\pm$ 2973~G. For comparison the derived field strength for the field star
amounts to $-438\pm 318/819$~G (68.3\%/99\% confidence range). 
Thus we conclude that GD\,40 does not posses a magnetic field with an 
upper limit of about 4000 G. 

Fig.~\ref{gd_pol} shows the observed circular polarization spectra of GD\,40 
at the H$\alpha$ line and the two \ion{He}{i} lines together with the respective 
polarization spectra with the magnetic field strengths predicted by 
the weak-field approximation and the maximum field strengths within the 68.3\% and 
99\% confidence ranges. 

\subsection{L745-46A}

\begin{figure}[htbp]
\includegraphics[width=0.45\textwidth]{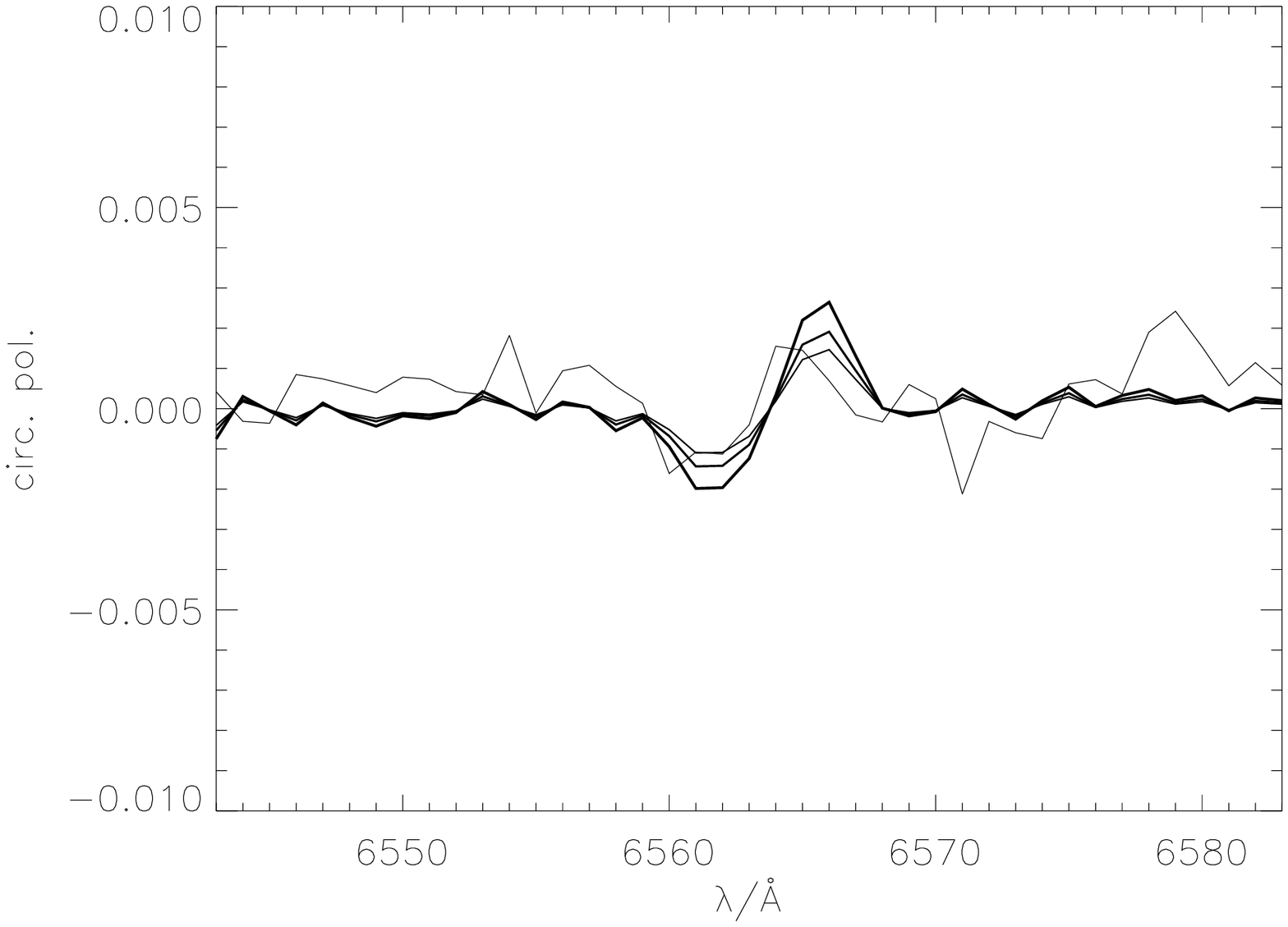}
\caption[]{Observed (gray) circular polarization spectrum of 
  L745-46A at H$\alpha$ 
  together with the polarization spectra (bold) determined by the  weak-field 
  approximation for field strengths of $-6900$~G, $-9000$~G, and $-12450$~G. 
  The first field strength denotes the best fit value, the second and 
  third value the 
  the best fit value plus the 68.3\%  and 99\% confidence range, respectively} 
\label{lpol}    
\end{figure}

Fig.~\ref{lpol} shows the circular polarization at 
H$\alpha$, which is the only visible line in the observed spectral range, 
together with the polarization spectrum for the best-fit magnetic 
field strength of $-6900$~G. The 68.3\% and 99\% confidence ranges amount to
$\pm 2100$~G and $\pm 5550$~G. Again the polarization spectra with the
maximum value from this field strength and the 68.3\% and 99\% confidence range are
also plotted. Although the field strength clearly exceeds the statistical 
1$\sigma$ error one has to be cautious, because the result solely 
depends on the vicinity of the H$\alpha$ line. However, if this field could 
be confirmed, it would be strong enough to prevent hydrogen from accreting 
according to the prediction from the propeller mechanism.  

\section{Conclusion}
We have obtained high signal-to-noise circular polarization spectra of the 
helium-rich white dwarfs GD\,40 and LP745-46A. With errors as low as 
$\pm$0.5\% and $\pm$0.2\% these spectra allow for the first time to
search for the weak magnetic fields necessary to drive the so called propeller
mechanism which is assumed to be responsible for the low observed 
hydrogen-to-metal ratios found in some cool helium-rich white dwarfs.
   
For L745-46A the predicted magnetic field strength from the weak field
approximation amounts to $-6900$~G with a formal $1\sigma$ error of 
2100~G and a 99\% confidence range of $\pm$5550~G. This magnetic field 
strength exceeds   
the minimum field strength of 3000~G necessary for the propeller 
mechanism to work. However, one should keep in mind, that this result is 
based on the H$\alpha$ line only. If the detection is confirmed, this means 
for the first time an indication that magnetic fields may play a role in
the accretion of hydrogen and metals onto cool helium-rich white dwarfs. 

However, we could not detect signatures of a magnetic field in the circular
polarization or flux spectrum of GD\,40. This does not necessarily 
mean, that 
no magnetic field is present, because if we are looking on the magnetic 
equator of a magnetic dipole ($i=90^\circ$), the components of the magnetic 
field along the line of sight completely cancel and no circular polarization 
can be detected. According to Wesemael \& Truran (1982) a field 
strength of about $144000$~G is
needed for GD\,40 to let the propeller mechanism work. If we now assume that
we could detect a field strength of 4000~G (derived upper limit field 
strength from the fit to all three lines with a confidence of 99\%) we 
could estimate an inclination angle of greater than 84$^\circ$ for which 
the longitudinal field strength becomes to low to be detected (Brown et 
al. \cite{brown77}). Therefore the chance to miss a magnetic field of 
144000~G is only about 6\%.

From our observations we must therefore conclude that at least for GD\,40 
other mechanisms than magnetic fields prevent hydrogen from accreting. 
Sion et al. (\cite{sion90}) argue that hydrogen deficient stars like 
GD\,40, which do not lie close to any known interstellar cloud (Aannestad \& 
Sion \cite{aannestad85}), have accreted, but not from interstellar clouds.
They proposed that white dwarfs might experience accretion in different 
environments with some white dwarfs being accreting material from 
interstellar clouds and some from volatile-free
asteroidal or dust material or chemically differentiated planetary
material.  
Accretion of comets from an Oort-like cloud as proposed by Alcock et al. 
(\cite{alcock86}), however, 
failed to explain the observed Ca/H ratios and is ruled out by Zuckerman
et al. (\cite{zuckerman03}) as a source of metals at least for DAZ 
white dwarfs. In view of the suggestion that the DAZ phenomenon
might be associated with the existence of a nearby unseen companion (Zuckerman
\& Reid \cite{zuckerman98}), Zuckerman et al. (\cite{zuckerman03}) 
also speculate about Jupiter-size substellar 
companions of DA and DB white dwarfs, that might generate detectable 
metal-line absorption. One problem with this model is lifting material 
off the secondary. Another problem might be the lack of observed eclipses, which 
would be expected, if a substantial number of all DAs and DBs are orbited by 
such substellar
companions with semi-major axis of a few solar radii, which are typical for red
dwarf/white dwarf pairs. A dusty circumstellar disk might be also the reason 
for a hydrogen depleted environment as mentioned in Aannestad et al. 
(\cite{aannestad93}). The maintenance of such a disk over the long 
lifetime of a cool white dwarf, however, would be difficult to understand.

From the determination of hydrogen and metal abundances in helium-rich white
dwarfs one must conclude that predominantly metals are accreted. However, the
reasons for this selective accretion process are not yet really understood. It
seems that also other mechanism than magnetic fields must be considered.

\begin{acknowledgements}
We thank the ESO staff on Cerro Paranal
for observing GD\,40 and L745-46A with the VLT in service mode.
Work on magnetic white dwarfs in Kiel was supported by the DFG under 
KO-738/7-1. In T\"ubingen work is supported by the DLR under 50 OR 0201. 
\end{acknowledgements}

{}

\end{document}